\documentclass[%
 preprint,
 amsmath,amssymb,
 aps,
 prl,
]{revtex4-2}

\usepackage{graphicx}
\usepackage{dcolumn}
\usepackage{bm}

\begin{document}

\preprint{APS/123-QED}

\title{Ultra Long Turbulent Eddies, Magnetic Topology, and the Triggering of Internal Transport Barriers in Tokamaks}

\author{Arnas Vol\v{c}okas}
 \email{Arnas.Volcokas@epfl.ch}
\author{Justin Ball}%
\author{Stephan Brunner}%
\affiliation{%
 Ecole Polytechnique F\'{e}d\'{e}rale de Lausanne (EPFL), Swiss Plasma Center (SPC), CH-1015 Lausanne, Switzerland 
}%

\date{\today}

\begin{abstract}
Local nonlinear gyrokinetic simulations of tokamak plasmas demonstrate that turbulent eddies can extend along magnetic field lines for hundreds of poloidal turns when the magnetic shear is very small. By accurately modeling different field line topologies (e.g. low-order, almost rational, or irrational safety factor), we show that the parallel self-interaction of such ``ultra long’’ eddies can dramatically reduce heat transport. This reveals novel strategies to improve confinement, constitutes experimentally testable predictions, and illuminates past observations of internal transport barriers.

\end{abstract}

\maketitle

\paragraph*{Introduction. ---}

Understanding and reducing the turbulence that transports energy out of the plasma is crucial to designing an economically competitive magnetic confinement fusion power plant \cite{Freidberg2007}. In this Letter, we will present high-fidelity local (flux tube) gyrokinetic simulations of tokamak plasmas demonstrating that nonlinear turbulent eddies become much longer than previously appreciated at very low magnetic shear $\hat{s} \approx 0$ (i.e. weak variation of the field line pitch between magnetic surfaces). While eddies are typically assumed to have lengths comparable to the connection length \cite{Barnes2011, Ottaviani1997} (i.e. the distance along the magnetic field between the outboard and inboard midplanes of the torus), when $\hat{s} \approx 0$ we show that they can be hundreds of times longer. This can be seen as a nonlinear analogue of ``giant electron tails'' --- extended parallel structures observed in linear modes that result from kinetic electron physics \cite{Hallatschek2005, Hardman2022}.

When eddies become hundreds of poloidal turns long, they can wrap around the device many times and ``bite their own tail.'' As we will see, the resulting parallel self-interaction can dramatically alter the transport properties of the turbulence. Hence, to accurately simulate real experiments with low $\hat{s}$, it becomes necessary to faithfully model the correct field line topology. This has rarely been studied. Even the original paper formulating the flux tube \cite{Beer1995} recommended always increasing the length of the simulation domain in order to entirely eliminate self-interaction. More recent works have investigated the impact of field line topology, but for standard values of shear $\hat{s} \sim 1$ \cite{Waltz2010, Pueschel2013, Dominski2015, Weikl2018, Ball2020, Ajay2020, Ajay2021, Ajay2022}. While some global gyrokinetic simulations can correctly model the topology, due to their high computational cost, there have been few studies that included both $\hat{s} \approx 0$ and kinetic electrons \cite{Garbet2010}. An important exception is \cite{Waltz2006}, which investigated Internal Transport Barriers (ITBs). ITBs are an improved confinement regime characterized by a localized steepening of the pressure profile in the core \cite{Wolf2003, Connor2004a, Ida2018}. Importantly, experimental observations have revealed that ITBs are enabled by low $\hat{s}$ and preferentially form near surfaces with rational values of the safety factor $q$ (i.e. the number of toroidal turns field lines make around the torus per poloidal turn) \cite{Joffrin2002, Eriksson2002}. Accordingly, \cite{Waltz2006} found large corrugations in temperature, density, and flow profiles associated with $\hat{s} \approx 0$ rational surfaces, but with limited insight into the physics at play.

In this Letter, we employ gyrokinetics, a high fidelity kinetic model of magnetized plasma turbulence \cite{Catto1978, Frieman1982, Brizard2007, Abel2013}, and solve the coupled system of gyrokinetic Vlasov and Maxwell's equations numerically using the spectral code GENE \cite{Jenko2000}. The results that follow were obtained with collisionless, electrostatic simulations and require a full kinetic treatment of both ions and electrons. We specify the equilibrium with Miller geometry \cite{Miller1998} and use a local domain \cite{Beer1995}. A local treatment takes advantage of the anisotropic nature of the fluctuations and constructs a long, slender domain along magnetic field lines where all equilibrium quantities are Taylor expanded around a magnetic surface of interest and only vary along the magnetic field.

We will study standard Ion Temperature Gradient (ITG)-driven turbulence at zero magnetic shear $\hat{s} = 0$ using two cases: the standard Cyclone Base Case (CBC) \cite{Dimits2000} and a ``pure ITG'' (pITG) case. The latter is simply the former without the density and {\it electron} temperature gradients, so that the ion temperature gradient is the only source of free energy driving turbulence. We choose these two cases as we will find them to be dominated by different ITG instability branches. We will employ both linear and nonlinear local gyrokinetic simulations and vary two parameters, $N_{pol}$ and $\Delta y$, to understand the impact of topology. The parameter $N_{pol}$ is the length of the simulation domain in number of poloidal turns, which controls the distance that an eddy must stretch before it can return to its starting parallel location. The parameter $\Delta y$ is the shift in the binormal direction that all field lines experience as they pass through the parallel boundary of the domain (i.e. in the direction perpendicular to the field line, but within the magnetic surface). Mathematically, this shift enters the parallel boundary condition as \cite{Ball2020} 
\begin{equation}
    \label{eq:ParallelBoundary}
    \delta A(x, y + (\Delta y + N_{*} L_{y}), z + 2\pi N_{pol}) = \delta A(x,y,z) \; ; \Delta y < L_{y} ,
\end{equation}
where $\delta A$ represents any turbulent quantity, $L_{y}$ is the binormal domain width, $x$ and $y$ are the minor radial and the binormal coordinates respectively, and $z$ is the poloidal angle (i.e. the parallel coordinate). Statistically motivated periodic boundary conditions are used in the x and y directions, so applying binormal periodicity removes $N_{*}L_{y}$ (given that $N_{*} \in \mathbb{N}$). Thus, $\Delta y = 0$ creates field lines that all exactly close on themselves after one time through the domain and, for example, setting $\Delta y = L_{y}/2$ will let field lines pass through the parallel boundary twice before closing. At finite shear $\Delta y$ corresponds to a simple radial shift of the lowest order rational surfaces, so without loss of generality $\Delta y$ can be set to zero, but this is no longer true when $\hat{s} = 0$ \cite{Beer1995}. Accordingly, we have implemented the shifted parallel boundary condition in GENE and benchmarked the changes against analytical ITG and parallel velocity gradient (PVG) instabilities in the cold ion limit \cite{Ball2019, Volcokas2022}. Including $\Delta y$ is essential to properly model physically possible magnetic topologies.  

While appropriately choosing $N_{pol}$ and $\Delta y$ given an arbitrary physical equilibrium is quite complicated \cite{Volcokas2022}, many cases of practical importance are straightforward. For example, a field line on a $q = 3$ surface exactly closes on itself after one poloidal turn, so one should take $N_{pol} = 1$ and $\Delta y = 0$. Similarly, a $q = 5/2$ field line closes after two turns, meaning one can take $N_{pol} = 2$ and $\Delta y = 0$. Finite $\Delta y$ is needed for surfaces that are very close to, but not exactly rational. For example, since eddies are $\sim 10 \rho_{i}$ wide in the binormal direction, if a field line comes within four ion gyroradii $\rho_{i}$ of closing on itself after one poloidal turn, we should set $N_{pol} = 1$ and $\Delta y = 4 \rho_{i}$. This ensures that we capture the complexity of an eddy partially (but not exactly) biting its own tail. Given the scale separation of gyrokinetics, a field line that comes within a few ion gyroradii of being integer would correspond to a physical value of $q$ that is asymptotically close to, but not actually a rational, e.g. $q = 3 + O ( \rho_{i}/a )$ where $a$ is the device minor radius.

\paragraph*{Linear simulations. ---}

In linear simulations with $N_{pol}=1$ and $\Delta y = 0$ we observe unique behavior as the magnetic shear drops below $|\hat{s}| \lesssim 0.1$. The ballooning structure of eigenmodes become broader and toroidal ITG (tITG) weakens relative to slab ITG (sITG), causing a transition in the dominant linear mode in both the CBC and pITG cases. We identify the tITG $\rightarrow$ sITG mode transition based on qualitative agreement with previous analytical results \cite{Romanelli1993, Connor2004}, a large discontinuous increase in the real frequency, and a sharp broadening of the eigenmode in ballooning space. This broadening at low magnetic shear suggests self-interaction is becoming increasingly important.

\begin{figure}[b]
\includegraphics{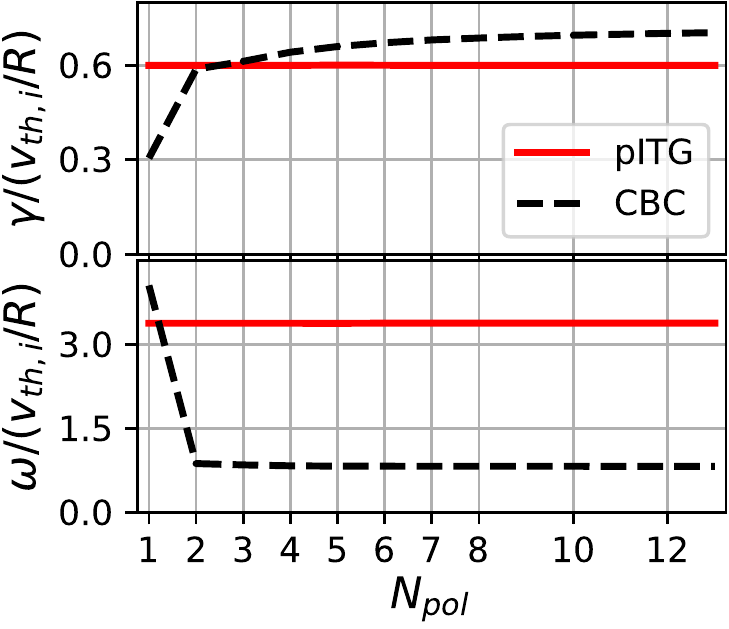}
\caption{\label{fig:PRL_LinModeFreqVSNpol} The linear mode growth rate (top) and real frequency (bottom) with $N_{pol}$ for CBC and pITG parameters using $\hat{s}=0$, $\Delta y = 0$, the binormal wavenumber $k_{y} \rho_{i} = 0.45$, and a radial wavenumber chosen to capture the fastest growing mode. Here $v_{th,i}$ is the ion thermal velocity and $R$ is the major radius.}
\end{figure}

Accordingly, Fig. \ref{fig:PRL_LinModeFreqVSNpol} shows a scan in $N_{pol}$ at $\Delta y = 0$ and $\hat{s} = 0$. Physically, this corresponds to studying how linear stability is affected by changing the order of the rational surfaces (e.g. integer, half-integer, third-integer). We see that the sITG mode is independent of the order of the rational surface, whereas the tITG mode is stabilized by low order rational surfaces. In a long domain, sITG modes are dominant for pITG gradients and tITG modes for CBC gradients. However, as the domain is shortened the tITG instability is stabilized and transitions to sITG at $N_{pol}=1$ for CBC as well. 

\begin{figure}[b]
\includegraphics{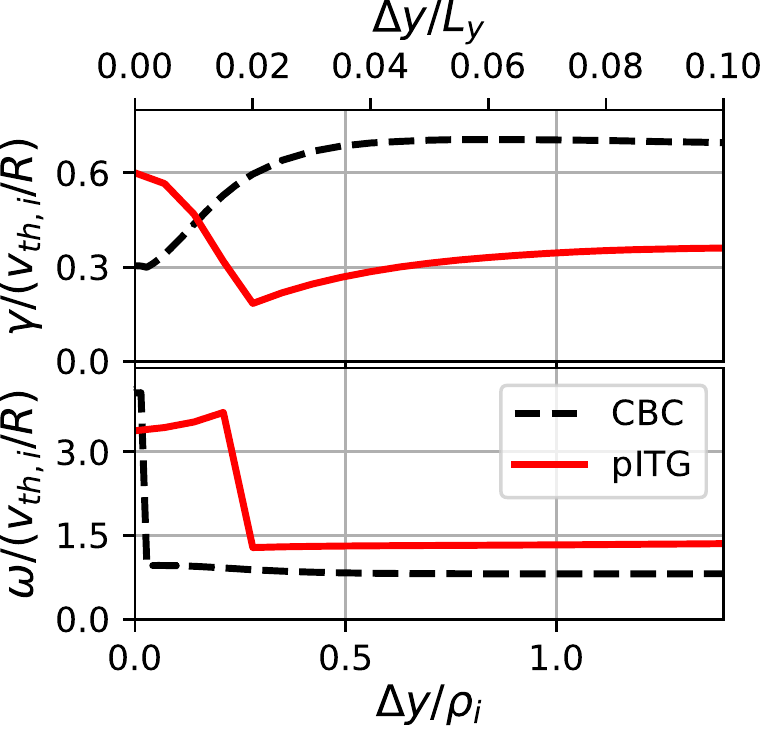}
\caption{\label{fig:PRL_LinModeFreqVSeta} The linear mode growth rate (top) and real frequency (bottom) with the binormal shift $\Delta y$ for CBC and pITG parameters using $\hat{s}=0$, $N_{pol}=1$, the binormal wavenumber $k_{y} \rho_{i} = 0.45$, and a radial wavenumber chosen to capture the fastest growing mode.}
\end{figure}

Figure \ref{fig:PRL_LinModeFreqVSeta} shows how $\Delta y$ changes the linear drive when $\hat{s}=0$ and $N_{pol}=1$, indicating the impact of magnetic field lines not exactly closing on themselves. We see that small changes of order $\Delta y \approx \rho_{i}$ can cause the dominant instability to transition between sITG and tITG and change the growth rate by a factor of 2 (finite $\Delta y$ increases the growth rate for CBC and decreases it for pITG). Though not shown here, in some cases changing $\Delta y$ can even control if the mode is linearly unstable or not. We also see that the growth rate for sITG peaks around $\Delta y = 0$, i.e. when magnetic field lines exactly close, whereas the tITG growth rate is maximum at $\Delta y \approx \rho_{i}$. Thus, we can expect that sITG will tend to be strongest exactly on rational surfaces and tITG away from them.

Finally, using Floquet-Bloch theory \cite{Floquet1883, Bloch1929}, the linear growth rates from a $\Delta y$ scan with $N_{pol}=1$ (e.g. Fig. \ref{fig:PRL_LinModeFreqVSeta}) can be used to analytically calculate growth rates for any value of $N_{pol}>1$ and $\Delta y$ \cite{Volcokas2022}. This is because each poloidal turn is identical due to axisymmetry. For instance, a simulation with $N_{pol} = 4$ and a given $\Delta y |_{N_{pol}=4}$ is equivalent to 4 identical linked $N_{pol}=1$ simulations, each with a binormal shift $\Delta y |_{N_{pol}=1}$ that, when repeated 4 times, corresponds to $\Delta y |_{N_{pol}=4}$. Thus, the allowed $\Delta y|_{N_{pol}=1}$ values are $\Delta y|_{N_{pol}=1} = ( \Delta y |_{N_{pol}=4} + p L_{y} ) / 4$, where $p \in \mathbb{N}$ is a free parameter. Each of these allowed $\Delta y|_{N_{pol}=1}$ has a different growth rate, the fastest growing of which will dominate and set the growth rate of the $N_{pol}=4$ simulation. Generalizing this to domains longer than $N_{pol} = 4$, we see that as $N_{pol} \rightarrow \infty$ the binormal shift of the domain ceases to matter and the free parameter $p$ always enables the simulation to find the maximum growth rate from the $\Delta y$ scan with $N_{pol} = 1$.

\paragraph*{Nonlinear simulations. ---}

\begin{figure}[b]
\includegraphics{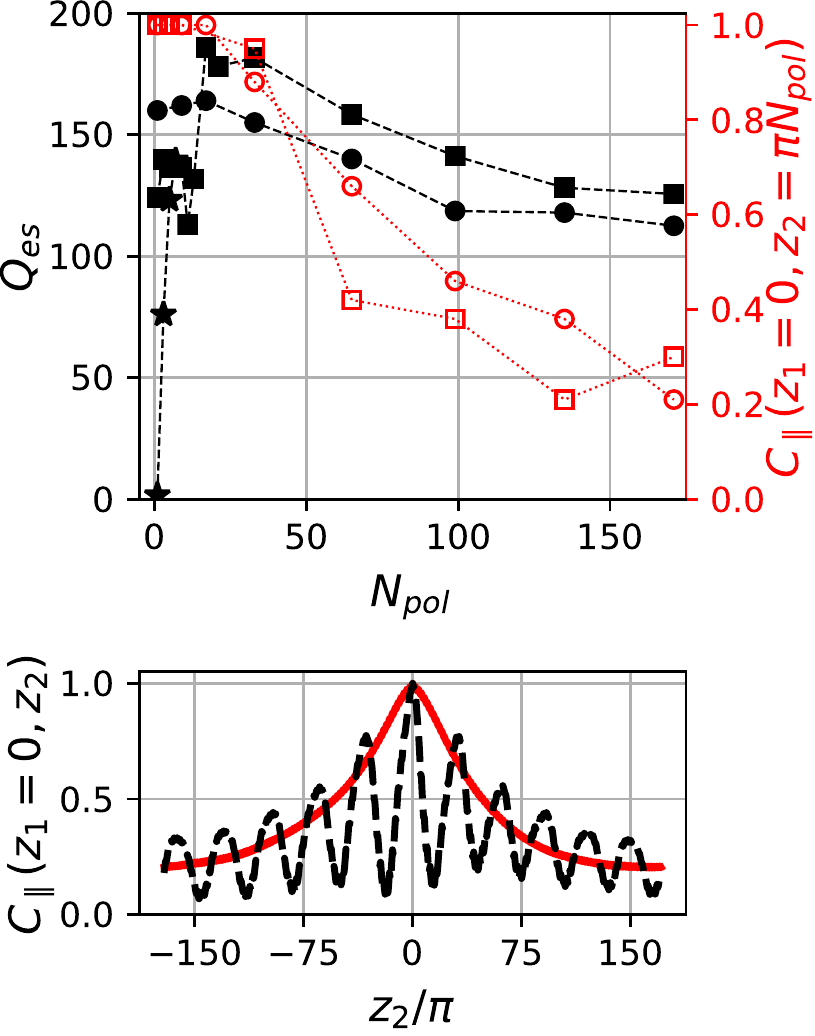}
\caption{\label{fig:PRL_CBCpITG_s0_NpolScan} (Top) The total electrostatic heat flux and parallel correlation with the parallel domain length $N_{pol}$ for CBC with $\Delta y =0$ (squares), pITG with $\Delta y =0$ (circles) and pITG with $\Delta y \approx 0.6 \rho_{i}$ (stars) parameters. The statistical error bars are within the marker size. (Bottom) The parallel correlation within the longest simulation domains for CBC (dashed black) and pITG (solid red) parameters.}
\end{figure}

Analogously to the linear study of Fig. \ref{fig:PRL_LinModeFreqVSNpol}, we carried out a nonlinear scan in $N_{pol}$ with $\hat{s} = 0$. Figure \ref{fig:PRL_CBCpITG_s0_NpolScan} shows how the total electrostatic heat flux $Q_{es}$ and $C_{||}(z_{1}, z_{2})$ vary with the parallel domain length, where $C_{||}(z_{1}, z_{2})$ is the two-point parallel correlation function \cite{Ball2020} of the non-zonal electrostatic potential averaged over time, $x$, and $y$. We see that the heat flux requires $N_{pol} > 100$ for convergence and parallel locations more than 50 poloidal turns apart remain highly correlated --- direct evidence of ultra long turbulent eddies. In fact, we believe their length is only limited by critical balance --- the distance that electrons can travel along the field line within the lifetime of a turbulent eddy \cite{Barnes2011}. Accordingly, simulations confirm that if the electron thermal velocity is artificially reduced (by increasing their mass), the parallel correlation length shrinks proportionally. In contrast, simulations with an adiabatic approximation for electrons exhibit parallel correlation spanning no more than a couple poloidal turns, reflecting the distance an ion can travel in an eddy's lifetime. Thus, fully resolving kinetic electron effects (in particular with the physical $m_{e}/m_{i}$ mass ratio) is crucial to properly model low magnetic shear surfaces.  

At small $N_{pol}$ we expect the heat flux in Fig. \ref{fig:PRL_CBCpITG_s0_NpolScan} to follow the linear trends of Fig. \ref{fig:PRL_LinModeFreqVSNpol}. Indeed, the heat flux has little variation for the pITG case as expected. However, for CBC, while the small jump in the heat flux at $N_{pol}=3$ could be consistent with the increase in the linear growth rate, a larger absolute change was expected. Hence, another $N_{pol}$ scan was performed with $\Delta y \approx 0.6 \rho_{i}$ for the pITG case, which shows a sharp increase with $N_{pol}$ as is expected from linear results. At large $N_{pol} > 40$, both parameter sets exhibit a similar decrease in the heat flux and parallel correlation, suggesting a common mechanism. This appears to be a result of the parallel domain being long enough to accommodate multiple eddies, instead of just one. When there is just a single correlated eddy, it can organize internally to maximize transport. However, when there is more than one eddy along the parallel direction they can organize within themselves, but there will always be transition regions where the different eddies meet and the radial transport is reduced.

Additionally, in the CBC simulations we observed long parallel waves that appeared when the domain exceeded $N_{pol} \approx 20$. As with the eddy length, their wavelength is proportional to the thermal electron velocity. These waves cause the oscillations in the parallel correlation profile shown in Fig. \ref{fig:PRL_CBCpITG_s0_NpolScan} (bottom). Their appearance is associated with the $\sim 30 \%$ increase in the particle and heat transport as shown in Fig. \ref{fig:PRL_CBCpITG_s0_NpolScan} (top). Thus, they seem important, but a detailed study of them is left for future work.

\begin{figure}[b]
\includegraphics{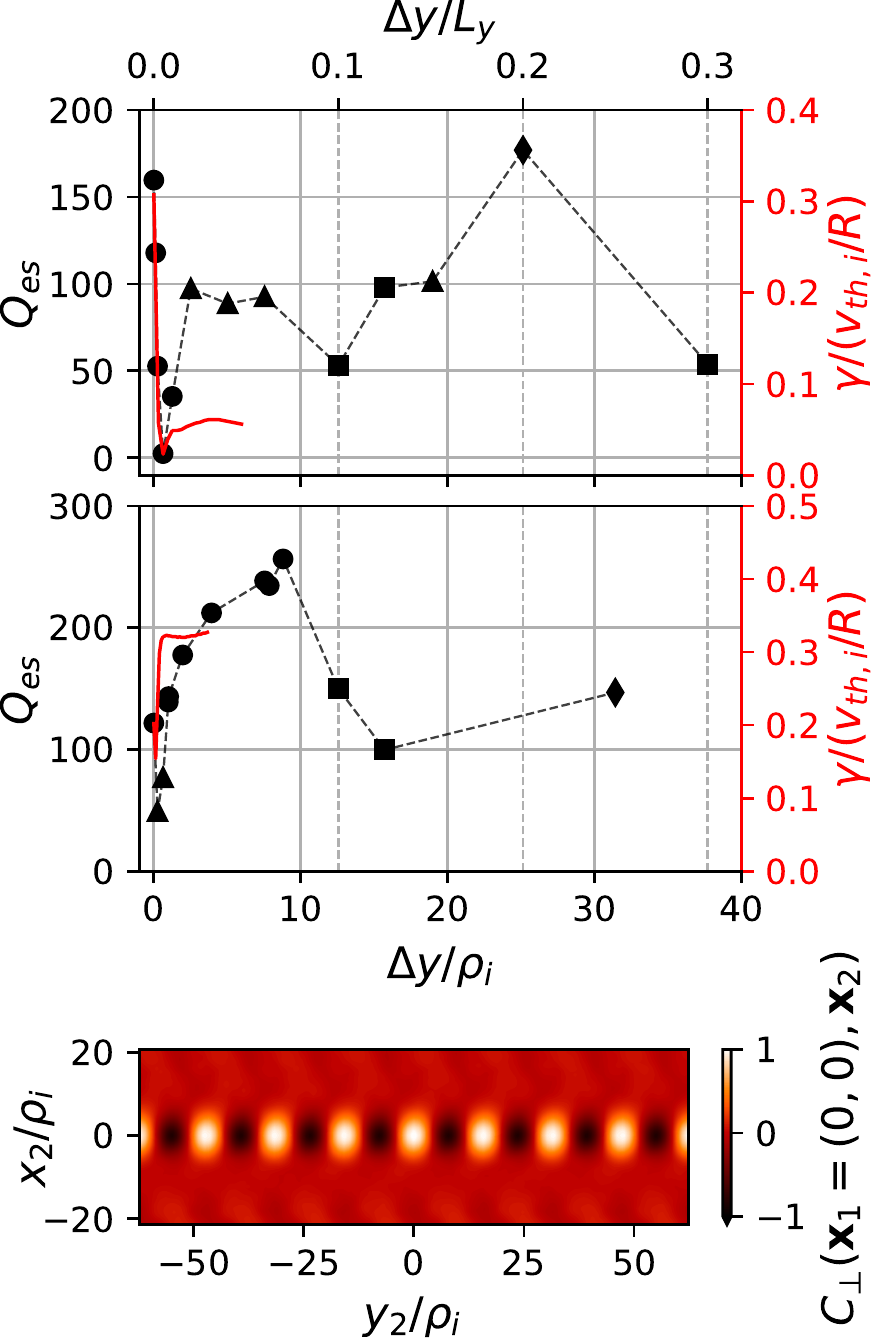}
\caption{\label{fig:PRL_CBCpITGetaScan} The total electrostatic heat flux with the binormal shift $\Delta y$ for pITG (top) and CBC (middle) parameters. The marker type is explained in the text and the solid red line shows linear growth rate at the binormal wavenumber where the nonlinear spectra peaks. The statistical error bars are within marker size. (Bottom) Contours of perpendicular two-point correlation at $z=0$ between the center $\mathbf{x}_{1}=(0,0)$ and all other points at $\mathbf{x}_{2}=(x_{2}, y_{2})$ for $\Delta y= L_{y}/8$ illustrating poloidal eddy squeezing.}
\end{figure}

Lastly, we scanned the binormal shift $\Delta y$ for domains with $N_{pol} = 1$ and $\hat{s}=0$, which is equivalent to considering near integer $q$. We observe dramatic, fine-scale variation in the heat flux with $\Delta y$ as seen in Fig. \ref{fig:PRL_CBCpITGetaScan}. While this is surprising, independent evidence of this can be found in \cite{StOnge2022}. We see that low but finite $\Delta y \leq \rho_{i}$ (circle markers) can double the heat flux or completely stabilize turbulence, which can be explained by the variation of the linear drive (thin red line). At somewhat larger values of $\Delta y$ (triangle markers), the heat flux time traces become bursty, alternating between two distinct values of heat flux. During the peak of the bursts for pITG case (top panel), turbulence shows strong parallel correlation across the domain $C_{||}(z_{1} = 0, z_{2} = \pi) \approx 0.8$, indicating that the turbulence has exactly bitten its tail. During the troughs of the bursts, the parallel correlation $C_{||}(z_{1} = 0, z_{2}) \approx 0.4$ is much lower, suggesting that the eddies don't exactly bite their own tails. This trend flips for CBC simulations (as expected from Fig. \ref{fig:PRL_LinModeFreqVSeta})--- turbulence shows weak parallel correlation during the peak of the burst and high correlation in the troughs. Thus, at these values of $\Delta y$ (triangle markers) the eddies stochastically connect/disconnect from themselves and alternate between the two states due to chaotic nature of the system. At larger values of $\Delta y$ corresponding to high-order rational surfaces (square markers), the eddies pass through the domain several times and then exactly bite their own tails. Since ultra long eddies can span hundreds of poloidal turns, on some high-order rational surfaces a single eddy can cover the full flux surface, even in a real device. To study this, we can set $L_{y}$ to correspond to the full flux surface of a given device. For example, on the eighth-order rational surface shown in Fig. \ref{fig:PRL_CBCpITGetaScan} (bottom), we see that a single eddy tiles the entire surface. More important, this eight-fold periodicity does {\it not} give the eddy enough space to be its preferred size to maximise transport. Instead the lack of space in the poloidal direction ``squeezes'' the eddy, restricting its size through {\it perpendicular} self-interaction and reduces its ability to transport energy. Such poloidal eddy squeezing presents a novel strategy for improving confinement. This is accomplished by tailoring the safety factor profile to have particular high-order rational values where $\hat{s} = 0$. It is attractive because, unlike integer surfaces, high-order rational surfaces do not enable magnetohydrodynamic (MHD) instabilities \cite{Wesson2004}. Lastly, at particular high-order rational values of $\Delta y$ (diamond markers) it is also possible to flip this effect and somewhat ``stretch'' eddies, thereby slightly increasing turbulent transport. 

\paragraph*{Conclusions. ---}

We have observed turbulent eddies that span hundreds of poloidal turns in gyrokinetic simulations of tokamaks with low magnetic shear $\hat{s} \approx 0$. Such ultra long eddies mean that the topology of the magnetic field becomes very important. A small binormal shift of $\Delta y \approx \rho_{i}$ at the parallel boundary (corresponding to a small change in the safety factor away from being rational) was found to either completely stabilise turbulence or increase the heat flux by a factor of two. This demonstrates the importance of faithfully constructing the topology of the device being modeled. Moreover, it may explain the experimental observation that ITBs are easier to trigger in the vicinity of rational surfaces. Additionally, we have seen that an individual ultra long eddy may have the ability to entirely cover a flux surface, even in large devices. This enables a novel strategy to improve confinement --- by carefully selecting a high-order rational value of the safety factor, the eddy can box itself in and squeeze itself to a smaller perpendicular size. Several important considerations like plasma shaping (which can cause strong {\it local} magnetic shear), collisions (which may shorten the parallel correlation length), and electromagnetic effects (which allow the turbulence itself to modify the magnetic field geometry) are left for future work. Additionally, stellarators (which commonly have very weak global magnetic shear) have exhibited \textit{improved} confinement around rational magnetic surfaces \cite{Brakel2002}, but are also left for future study.

\begin{acknowledgments}
The authors would like to thank Oleg Krutkin, Alessandro Geraldini and Antoine Hoffmann for useful discussions pertaining to this work.
This work has been carried out within the framework of the EUROfusion Consortium, funded by the European Union via the Euratom Research and Training Programme (Grant Agreement No 101052200 --- EUROfusion). Views and opinions expressed are however those of the author(s) only and do not necessarily reflect those of the European Union or the European Commission. Neither the European Union nor the European Commission can be held responsible for them.
We acknowledge the CINECA award under the ISCRA initiative, for the availability of high performance computing resources and support. 
This work was supported by a grant from the Swiss National Supercomputing Centre (CSCS) under project ID s1050 and s1097. 
This work was supported in part by the Swiss National Science Foundation.

\end{acknowledgments}

\bibliography{prlref}

\begin{thebibliography}{37}%
\makeatletter
\providecommand \@ifxundefined [1]{%
 \@ifx{#1\undefined}
}%
\providecommand \@ifnum [1]{%
 \ifnum #1\expandafter \@firstoftwo
 \else \expandafter \@secondoftwo
 \fi
}%
\providecommand \@ifx [1]{%
 \ifx #1\expandafter \@firstoftwo
 \else \expandafter \@secondoftwo
 \fi
}%
\providecommand \natexlab [1]{#1}%
\providecommand \enquote  [1]{``#1''}%
\providecommand \bibnamefont  [1]{#1}%
\providecommand \bibfnamefont [1]{#1}%
\providecommand \citenamefont [1]{#1}%
\providecommand \href@noop [0]{\@secondoftwo}%
\providecommand \href [0]{\begingroup \@sanitize@url \@href}%
\providecommand \@href[1]{\@@startlink{#1}\@@href}%
\providecommand \@@href[1]{\endgroup#1\@@endlink}%
\providecommand \@sanitize@url [0]{\catcode `\\12\catcode `\$12\catcode
  `\&12\catcode `\#12\catcode `\^12\catcode `\_12\catcode `\%12\relax}%
\providecommand \@@startlink[1]{}%
\providecommand \@@endlink[0]{}%
\providecommand \url  [0]{\begingroup\@sanitize@url \@url }%
\providecommand \@url [1]{\endgroup\@href {#1}{\urlprefix }}%
\providecommand \urlprefix  [0]{URL }%
\providecommand \Eprint [0]{\href }%
\providecommand \doibase [0]{https://doi.org/}%
\providecommand \selectlanguage [0]{\@gobble}%
\providecommand \bibinfo  [0]{\@secondoftwo}%
\providecommand \bibfield  [0]{\@secondoftwo}%
\providecommand \translation [1]{[#1]}%
\providecommand \BibitemOpen [0]{}%
\providecommand \bibitemStop [0]{}%
\providecommand \bibitemNoStop [0]{.\EOS\space}%
\providecommand \EOS [0]{\spacefactor3000\relax}%
\providecommand \BibitemShut  [1]{\csname bibitem#1\endcsname}%
\let\auto@bib@innerbib\@empty
\bibitem [{\citenamefont {Freidberg}(2007)}]{Freidberg2007}%
  \BibitemOpen
  \bibfield  {author} {\bibinfo {author} {\bibfnamefont {J.~P.}\ \bibnamefont
  {Freidberg}},\ }\href {https://doi.org/10.1017/CBO9780511755705} {}\
  (\bibinfo  {publisher} {Cambridge University Press},\ \bibinfo {year}
  {2007})\BibitemShut {NoStop}%
\bibitem [{\citenamefont {Barnes}\ \emph {et~al.}(2011)\citenamefont {Barnes},
  \citenamefont {Parra},\ and\ \citenamefont {Schekochihin}}]{Barnes2011}%
  \BibitemOpen
  \bibfield  {author} {\bibinfo {author} {\bibfnamefont {M.}~\bibnamefont
  {Barnes}}, \bibinfo {author} {\bibfnamefont {F.~I.}\ \bibnamefont {Parra}},\
  and\ \bibinfo {author} {\bibfnamefont {A.~A.}\ \bibnamefont {Schekochihin}},\
  }\href@noop {} {\bibfield  {journal} {\bibinfo  {journal} {Phys. Rev. Lett.}\
  }\textbf {\bibinfo {volume} {107}},\ \bibinfo {pages} {1} (\bibinfo {year}
  {2011})}\BibitemShut {NoStop}%
\bibitem [{\citenamefont {Ottaviani}\ \emph {et~al.}(1997)\citenamefont
  {Ottaviani}, \citenamefont {Beer}, \citenamefont {Cowley}, \citenamefont
  {Horton},\ and\ \citenamefont {Krommes}}]{Ottaviani1997}%
  \BibitemOpen
  \bibfield  {author} {\bibinfo {author} {\bibfnamefont {M.}~\bibnamefont
  {Ottaviani}}, \bibinfo {author} {\bibfnamefont {M.~A.}\ \bibnamefont {Beer}},
  \bibinfo {author} {\bibfnamefont {S.~C.}\ \bibnamefont {Cowley}}, \bibinfo
  {author} {\bibfnamefont {W.}~\bibnamefont {Horton}},\ and\ \bibinfo {author}
  {\bibfnamefont {J.}~\bibnamefont {Krommes}},\ }\href@noop {} {\bibfield
  {journal} {\bibinfo  {journal} {Phys. Rep.}\ }\textbf {\bibinfo {volume}
  {283}},\ \bibinfo {pages} {121} (\bibinfo {year} {1997})}\BibitemShut
  {NoStop}%
\bibitem [{\citenamefont {Hallatschek}\ and\ \citenamefont
  {Dorland}(2005)}]{Hallatschek2005}%
  \BibitemOpen
  \bibfield  {author} {\bibinfo {author} {\bibfnamefont {K.}~\bibnamefont
  {Hallatschek}}\ and\ \bibinfo {author} {\bibfnamefont {W.}~\bibnamefont
  {Dorland}},\ }\href {https://doi.org/10.1103/PhysRevLett.95.055002}
  {\bibfield  {journal} {\bibinfo  {journal} {Phys. Rev. Lett.}\ }\textbf
  {\bibinfo {volume} {95}},\ \bibinfo {pages} {2} (\bibinfo {year}
  {2005})}\BibitemShut {NoStop}%
\bibitem [{\citenamefont {Hardman}\ \emph {et~al.}(2022)\citenamefont
  {Hardman}, \citenamefont {Parra}, \citenamefont {Chong}, \citenamefont
  {Adkins}, \citenamefont {Anastopoulos-Tzanis}, \citenamefont {Barnes},
  \citenamefont {Dickinson}, \citenamefont {Parisi},\ and\ \citenamefont
  {Wilson}}]{Hardman2022}%
  \BibitemOpen
  \bibfield  {author} {\bibinfo {author} {\bibfnamefont {M.}~\bibnamefont
  {Hardman}}, \bibinfo {author} {\bibfnamefont {F.}~\bibnamefont {Parra}},
  \bibinfo {author} {\bibfnamefont {C.}~\bibnamefont {Chong}}, \bibinfo
  {author} {\bibfnamefont {T.}~\bibnamefont {Adkins}}, \bibinfo {author}
  {\bibfnamefont {M.}~\bibnamefont {Anastopoulos-Tzanis}}, \bibinfo {author}
  {\bibfnamefont {M.}~\bibnamefont {Barnes}}, \bibinfo {author} {\bibfnamefont
  {D.}~\bibnamefont {Dickinson}}, \bibinfo {author} {\bibfnamefont
  {J.}~\bibnamefont {Parisi}},\ and\ \bibinfo {author} {\bibfnamefont
  {H.}~\bibnamefont {Wilson}},\ }\href@noop {} {\bibfield  {journal} {\bibinfo
  {journal} {Plasma Phys. Control. Fusion}\ }\textbf {\bibinfo {volume} {64}},\
  \bibinfo {pages} {055004} (\bibinfo {year} {2022})}\BibitemShut {NoStop}%
\bibitem [{\citenamefont {Beer}\ \emph {et~al.}(1995)\citenamefont {Beer},
  \citenamefont {Cowley},\ and\ \citenamefont {Hammett}}]{Beer1995}%
  \BibitemOpen
  \bibfield  {author} {\bibinfo {author} {\bibfnamefont {M.~A.}\ \bibnamefont
  {Beer}}, \bibinfo {author} {\bibfnamefont {S.~C.}\ \bibnamefont {Cowley}},\
  and\ \bibinfo {author} {\bibfnamefont {G.~W.}\ \bibnamefont {Hammett}},\
  }\href {https://doi.org/10.1063/1.871232} {\bibfield  {journal} {\bibinfo
  {journal} {Phys. Plasmas}\ }\textbf {\bibinfo {volume} {2}},\ \bibinfo
  {pages} {2687} (\bibinfo {year} {1995})}\BibitemShut {NoStop}%
\bibitem [{\citenamefont {Waltz}(2010)}]{Waltz2010}%
  \BibitemOpen
  \bibfield  {author} {\bibinfo {author} {\bibfnamefont {R.~E.}\ \bibnamefont
  {Waltz}},\ }\href@noop {} {\bibfield  {journal} {\bibinfo  {journal} {Phys.
  Plasmas}\ }\textbf {\bibinfo {volume} {17}},\ \bibinfo {pages} {072501}
  (\bibinfo {year} {2010})}\BibitemShut {NoStop}%
\bibitem [{\citenamefont {Pueschel}\ \emph {et~al.}(2013)\citenamefont
  {Pueschel}, \citenamefont {G{\"o}rler}, \citenamefont {Jenko}, \citenamefont
  {Hatch},\ and\ \citenamefont {Cianciara}}]{Pueschel2013}%
  \BibitemOpen
  \bibfield  {author} {\bibinfo {author} {\bibfnamefont {M.~J.}\ \bibnamefont
  {Pueschel}}, \bibinfo {author} {\bibfnamefont {T.}~\bibnamefont
  {G{\"o}rler}}, \bibinfo {author} {\bibfnamefont {F.}~\bibnamefont {Jenko}},
  \bibinfo {author} {\bibfnamefont {D.~R.}\ \bibnamefont {Hatch}},\ and\
  \bibinfo {author} {\bibfnamefont {A.~J.}\ \bibnamefont {Cianciara}},\
  }\href@noop {} {\bibfield  {journal} {\bibinfo  {journal} {Phys. Plasmas}\
  }\textbf {\bibinfo {volume} {20}},\ \bibinfo {pages} {102308} (\bibinfo
  {year} {2013})}\BibitemShut {NoStop}%
\bibitem [{\citenamefont {Dominski}\ \emph {et~al.}(2015)\citenamefont
  {Dominski}, \citenamefont {Brunner}, \citenamefont {G{\"o}rler},
  \citenamefont {Jenko}, \citenamefont {Told},\ and\ \citenamefont
  {Villard}}]{Dominski2015}%
  \BibitemOpen
  \bibfield  {author} {\bibinfo {author} {\bibfnamefont {J.}~\bibnamefont
  {Dominski}}, \bibinfo {author} {\bibfnamefont {S.}~\bibnamefont {Brunner}},
  \bibinfo {author} {\bibfnamefont {T.}~\bibnamefont {G{\"o}rler}}, \bibinfo
  {author} {\bibfnamefont {F.}~\bibnamefont {Jenko}}, \bibinfo {author}
  {\bibfnamefont {D.}~\bibnamefont {Told}},\ and\ \bibinfo {author}
  {\bibfnamefont {L.}~\bibnamefont {Villard}},\ }\href@noop {} {\bibfield
  {journal} {\bibinfo  {journal} {Phys. Plasmas}\ }\textbf {\bibinfo {volume}
  {22}},\ \bibinfo {pages} {062303} (\bibinfo {year} {2015})}\BibitemShut
  {NoStop}%
\bibitem [{\citenamefont {Weikl}\ \emph {et~al.}(2018)\citenamefont {Weikl},
  \citenamefont {Peeters}, \citenamefont {Rath}, \citenamefont {Seiferling},
  \citenamefont {Buchholz}, \citenamefont {Grosshauser},\ and\ \citenamefont
  {Strintzi}}]{Weikl2018}%
  \BibitemOpen
  \bibfield  {author} {\bibinfo {author} {\bibfnamefont {A.}~\bibnamefont
  {Weikl}}, \bibinfo {author} {\bibfnamefont {A.~G.}\ \bibnamefont {Peeters}},
  \bibinfo {author} {\bibfnamefont {F.}~\bibnamefont {Rath}}, \bibinfo {author}
  {\bibfnamefont {F.}~\bibnamefont {Seiferling}}, \bibinfo {author}
  {\bibfnamefont {R.}~\bibnamefont {Buchholz}}, \bibinfo {author}
  {\bibfnamefont {S.~R.}\ \bibnamefont {Grosshauser}},\ and\ \bibinfo {author}
  {\bibfnamefont {D.}~\bibnamefont {Strintzi}},\ }\href@noop {} {\bibfield
  {journal} {\bibinfo  {journal} {Phys. Plasmas}\ }\textbf {\bibinfo {volume}
  {25}} (\bibinfo {year} {2018})}\BibitemShut {NoStop}%
\bibitem [{\citenamefont {Ball}\ \emph {et~al.}(2020)\citenamefont {Ball},
  \citenamefont {Brunner},\ and\ \citenamefont {{Ajay C. J.}}}]{Ball2020}%
  \BibitemOpen
  \bibfield  {author} {\bibinfo {author} {\bibfnamefont {J.}~\bibnamefont
  {Ball}}, \bibinfo {author} {\bibfnamefont {S.}~\bibnamefont {Brunner}},\ and\
  \bibinfo {author} {\bibnamefont {{Ajay C. J.}}},\ }\href
  {https://doi.org/10.1017/S0022377820000197} {\bibfield  {journal} {\bibinfo
  {journal} {J. Plasma Phys.}\ }\textbf {\bibinfo {volume} {86}},\ \bibinfo
  {pages} {1} (\bibinfo {year} {2020})}\BibitemShut {NoStop}%
\bibitem [{\citenamefont {{Ajay C. J.}}\ \emph {et~al.}(2020)\citenamefont
  {{Ajay C. J.}}, \citenamefont {Brunner}, \citenamefont {Mcmillan},
  \citenamefont {Ball}, \citenamefont {Dominski},\ and\ \citenamefont
  {Merlo}}]{Ajay2020}%
  \BibitemOpen
  \bibfield  {author} {\bibinfo {author} {\bibnamefont {{Ajay C. J.}}},
  \bibinfo {author} {\bibfnamefont {S.}~\bibnamefont {Brunner}}, \bibinfo
  {author} {\bibfnamefont {B.}~\bibnamefont {Mcmillan}}, \bibinfo {author}
  {\bibfnamefont {J.}~\bibnamefont {Ball}}, \bibinfo {author} {\bibfnamefont
  {J.}~\bibnamefont {Dominski}},\ and\ \bibinfo {author} {\bibfnamefont
  {G.}~\bibnamefont {Merlo}},\ }\href@noop {} {\bibfield  {journal} {\bibinfo
  {journal} {J. Plasma Phys.}\ } (\bibinfo {year} {2020})},\ \Eprint
  {https://arxiv.org/abs/2005.02709} {2005.02709} \BibitemShut {NoStop}%
\bibitem [{\citenamefont {{Ajay C. J.}}\ \emph {et~al.}(2021)\citenamefont
  {{Ajay C. J.}}, \citenamefont {Brunner},\ and\ \citenamefont
  {Ball}}]{Ajay2021}%
  \BibitemOpen
  \bibfield  {author} {\bibinfo {author} {\bibnamefont {{Ajay C. J.}}},
  \bibinfo {author} {\bibfnamefont {S.}~\bibnamefont {Brunner}},\ and\ \bibinfo
  {author} {\bibfnamefont {J.}~\bibnamefont {Ball}},\ }\href@noop {} {\bibfield
   {journal} {\bibinfo  {journal} {Phys. Plasmas}\ }\textbf {\bibinfo {volume}
  {28}},\ \bibinfo {pages} {092303} (\bibinfo {year} {2021})}\BibitemShut
  {NoStop}%
\bibitem [{\citenamefont {{Ajay C. J.}}\ \emph {et~al.}(2022)\citenamefont
  {{Ajay C. J.}}, \citenamefont {McMillan},\ and\ \citenamefont
  {Pueschel}}]{Ajay2022}%
  \BibitemOpen
  \bibfield  {author} {\bibinfo {author} {\bibnamefont {{Ajay C. J.}}},
  \bibinfo {author} {\bibfnamefont {B.}~\bibnamefont {McMillan}},\ and\
  \bibinfo {author} {\bibfnamefont {M.~J.}\ \bibnamefont {Pueschel}},\
  }\href@noop {} {\bibfield  {journal} {\bibinfo  {journal} {arXiv 2207.09211}\
  } (\bibinfo {year} {2022})}\BibitemShut {NoStop}%
\bibitem [{\citenamefont {Garbet}\ \emph {et~al.}(2010)\citenamefont {Garbet},
  \citenamefont {Idomura}, \citenamefont {Villard},\ and\ \citenamefont
  {Watanabe}}]{Garbet2010}%
  \BibitemOpen
  \bibfield  {author} {\bibinfo {author} {\bibfnamefont {X.}~\bibnamefont
  {Garbet}}, \bibinfo {author} {\bibfnamefont {Y.}~\bibnamefont {Idomura}},
  \bibinfo {author} {\bibfnamefont {L.}~\bibnamefont {Villard}},\ and\ \bibinfo
  {author} {\bibfnamefont {T.~H.}\ \bibnamefont {Watanabe}},\ }\href@noop {}
  {\bibfield  {journal} {\bibinfo  {journal} {Nucl. Fusion}\ }\textbf {\bibinfo
  {volume} {50}} (\bibinfo {year} {2010})}\BibitemShut {NoStop}%
\bibitem [{\citenamefont {Waltz}\ \emph {et~al.}(2006)\citenamefont {Waltz},
  \citenamefont {Austin}, \citenamefont {Burrell},\ and\ \citenamefont
  {Candy}}]{Waltz2006}%
  \BibitemOpen
  \bibfield  {author} {\bibinfo {author} {\bibfnamefont {R.~E.}\ \bibnamefont
  {Waltz}}, \bibinfo {author} {\bibfnamefont {M.~E.}\ \bibnamefont {Austin}},
  \bibinfo {author} {\bibfnamefont {K.~H.}\ \bibnamefont {Burrell}},\ and\
  \bibinfo {author} {\bibfnamefont {J.}~\bibnamefont {Candy}},\ }\href@noop {}
  {\bibfield  {journal} {\bibinfo  {journal} {Phys. Plasmas}\ }\textbf
  {\bibinfo {volume} {13}} (\bibinfo {year} {2006})}\BibitemShut {NoStop}%
\bibitem [{\citenamefont {Wolf}(2003)}]{Wolf2003}%
  \BibitemOpen
  \bibfield  {author} {\bibinfo {author} {\bibfnamefont {R.~C.}\ \bibnamefont
  {Wolf}},\ }\href@noop {} {\bibfield  {journal} {\bibinfo  {journal} {Plasma
  Phys. Control. Fusion}\ }\textbf {\bibinfo {volume} {45}} (\bibinfo {year}
  {2003})}\BibitemShut {NoStop}%
\bibitem [{\citenamefont {Connor}\ \emph {et~al.}(2004)\citenamefont {Connor}
  \emph {et~al.}}]{Connor2004a}%
  \BibitemOpen
  \bibfield  {author} {\bibinfo {author} {\bibfnamefont {J.~W.}\ \bibnamefont
  {Connor}} \emph {et~al.},\ }\href@noop {} {\bibfield  {journal} {\bibinfo
  {journal} {Nucl. Fusion}\ }\textbf {\bibinfo {volume} {44}} (\bibinfo {year}
  {2004})}\BibitemShut {NoStop}%
\bibitem [{\citenamefont {Ida}\ and\ \citenamefont {Fujita}(2018)}]{Ida2018}%
  \BibitemOpen
  \bibfield  {author} {\bibinfo {author} {\bibfnamefont {K.}~\bibnamefont
  {Ida}}\ and\ \bibinfo {author} {\bibfnamefont {T.}~\bibnamefont {Fujita}},\
  }\href@noop {} {\bibfield  {journal} {\bibinfo  {journal} {Plasma Phys.
  Control. Fusion}\ }\textbf {\bibinfo {volume} {60}} (\bibinfo {year}
  {2018})}\BibitemShut {NoStop}%
\bibitem [{\citenamefont {Joffrin}\ \emph {et~al.}(2002)\citenamefont
  {Joffrin}, \citenamefont {Gorini}, \citenamefont {Challis}, \citenamefont
  {Hawkes}, \citenamefont {Hender}, \citenamefont {Howell}, \citenamefont
  {Maget}, \citenamefont {Mantica}, \citenamefont {Mazon}, \citenamefont
  {Sharapov},\ and\ \citenamefont {Tresset}}]{Joffrin2002}%
  \BibitemOpen
  \bibfield  {author} {\bibinfo {author} {\bibfnamefont {E.}~\bibnamefont
  {Joffrin}}, \bibinfo {author} {\bibfnamefont {G.}~\bibnamefont {Gorini}},
  \bibinfo {author} {\bibfnamefont {C.~D.}\ \bibnamefont {Challis}}, \bibinfo
  {author} {\bibfnamefont {N.~C.}\ \bibnamefont {Hawkes}}, \bibinfo {author}
  {\bibfnamefont {T.~C.}\ \bibnamefont {Hender}}, \bibinfo {author}
  {\bibfnamefont {D.~F.}\ \bibnamefont {Howell}}, \bibinfo {author}
  {\bibfnamefont {P.}~\bibnamefont {Maget}}, \bibinfo {author} {\bibfnamefont
  {P.}~\bibnamefont {Mantica}}, \bibinfo {author} {\bibfnamefont
  {D.}~\bibnamefont {Mazon}}, \bibinfo {author} {\bibfnamefont {S.~E.}\
  \bibnamefont {Sharapov}},\ and\ \bibinfo {author} {\bibfnamefont
  {G.}~\bibnamefont {Tresset}},\ }\href
  {https://doi.org/10.1088/0741-3335/44/8/320} {\bibfield  {journal} {\bibinfo
  {journal} {Plasma Phys. Control. Fusion}\ }\textbf {\bibinfo {volume} {44}},\
  \bibinfo {pages} {1739} (\bibinfo {year} {2002})}\BibitemShut {NoStop}%
\bibitem [{\citenamefont {Eriksson}\ \emph {et~al.}(2002)\citenamefont
  {Eriksson} \emph {et~al.}}]{Eriksson2002}%
  \BibitemOpen
  \bibfield  {author} {\bibinfo {author} {\bibfnamefont {L.-G.}\ \bibnamefont
  {Eriksson}} \emph {et~al.},\ }\href@noop {} {\bibfield  {journal} {\bibinfo
  {journal} {Phys. Rev. Lett.}\ }\textbf {\bibinfo {volume} {88}},\ \bibinfo
  {pages} {145001} (\bibinfo {year} {2002})}\BibitemShut {NoStop}%
\bibitem [{\citenamefont {Catto}(1978)}]{Catto1978}%
  \BibitemOpen
  \bibfield  {author} {\bibinfo {author} {\bibfnamefont {P.~J.}\ \bibnamefont
  {Catto}},\ }\href@noop {} {\bibfield  {journal} {\bibinfo  {journal} {Plasma
  Phys.}\ }\textbf {\bibinfo {volume} {20}},\ \bibinfo {pages} {719} (\bibinfo
  {year} {1978})}\BibitemShut {NoStop}%
\bibitem [{\citenamefont {Frieman}\ and\ \citenamefont
  {Chen}(1982)}]{Frieman1982}%
  \BibitemOpen
  \bibfield  {author} {\bibinfo {author} {\bibfnamefont {E.~A.}\ \bibnamefont
  {Frieman}}\ and\ \bibinfo {author} {\bibfnamefont {L.}~\bibnamefont {Chen}},\
  }\href@noop {} {\bibfield  {journal} {\bibinfo  {journal} {Phys. Fluids}\
  }\textbf {\bibinfo {volume} {25}},\ \bibinfo {pages} {502} (\bibinfo {year}
  {1982})}\BibitemShut {NoStop}%
\bibitem [{\citenamefont {Brizard}\ and\ \citenamefont
  {Hahm}(2007)}]{Brizard2007}%
  \BibitemOpen
  \bibfield  {author} {\bibinfo {author} {\bibfnamefont {A.~J.}\ \bibnamefont
  {Brizard}}\ and\ \bibinfo {author} {\bibfnamefont {T.~S.}\ \bibnamefont
  {Hahm}},\ }\href {https://doi.org/10.1103/RevModPhys.79.421} {\bibfield
  {journal} {\bibinfo  {journal} {Rev. Mod. Phys.}\ }\textbf {\bibinfo {volume}
  {79}},\ \bibinfo {pages} {421} (\bibinfo {year} {2007})}\BibitemShut
  {NoStop}%
\bibitem [{\citenamefont {Abel}\ \emph {et~al.}(2013)\citenamefont {Abel},
  \citenamefont {Plunk}, \citenamefont {Wang}, \citenamefont {Barnes},
  \citenamefont {Cowley}, \citenamefont {Dorland},\ and\ \citenamefont
  {Schekochihin}}]{Abel2013}%
  \BibitemOpen
  \bibfield  {author} {\bibinfo {author} {\bibfnamefont {I.~G.}\ \bibnamefont
  {Abel}}, \bibinfo {author} {\bibfnamefont {G.~G.}\ \bibnamefont {Plunk}},
  \bibinfo {author} {\bibfnamefont {E.}~\bibnamefont {Wang}}, \bibinfo {author}
  {\bibfnamefont {M.}~\bibnamefont {Barnes}}, \bibinfo {author} {\bibfnamefont
  {S.~C.}\ \bibnamefont {Cowley}}, \bibinfo {author} {\bibfnamefont
  {W.}~\bibnamefont {Dorland}},\ and\ \bibinfo {author} {\bibfnamefont {A.~A.}\
  \bibnamefont {Schekochihin}},\ }\href@noop {} {\bibfield  {journal} {\bibinfo
   {journal} {Rep. Prog. Phys.}\ }\textbf {\bibinfo {volume} {76}},\ \bibinfo
  {pages} {116201} (\bibinfo {year} {2013})}\BibitemShut {NoStop}%
\bibitem [{\citenamefont {Jenko}\ \emph {et~al.}(2000)\citenamefont {Jenko},
  \citenamefont {Dorland}, \citenamefont {Kotschenreuther},\ and\ \citenamefont
  {Rogers}}]{Jenko2000}%
  \BibitemOpen
  \bibfield  {author} {\bibinfo {author} {\bibfnamefont {F.}~\bibnamefont
  {Jenko}}, \bibinfo {author} {\bibfnamefont {W.}~\bibnamefont {Dorland}},
  \bibinfo {author} {\bibfnamefont {M.}~\bibnamefont {Kotschenreuther}},\ and\
  \bibinfo {author} {\bibfnamefont {B.~N.}\ \bibnamefont {Rogers}},\ }\href
  {https://doi.org/10.1063/1.874014} {\bibfield  {journal} {\bibinfo  {journal}
  {Phys. Plasmas}\ }\textbf {\bibinfo {volume} {7}},\ \bibinfo {pages} {1904}
  (\bibinfo {year} {2000})}\BibitemShut {NoStop}%
\bibitem [{\citenamefont {Miller}\ \emph {et~al.}(1998)\citenamefont {Miller},
  \citenamefont {Chu}, \citenamefont {Greene}, \citenamefont {Lin-Liu},\ and\
  \citenamefont {Waltz}}]{Miller1998}%
  \BibitemOpen
  \bibfield  {author} {\bibinfo {author} {\bibfnamefont {R.~L.}\ \bibnamefont
  {Miller}}, \bibinfo {author} {\bibfnamefont {M.~S.}\ \bibnamefont {Chu}},
  \bibinfo {author} {\bibfnamefont {J.~M.}\ \bibnamefont {Greene}}, \bibinfo
  {author} {\bibfnamefont {Y.~R.}\ \bibnamefont {Lin-Liu}},\ and\ \bibinfo
  {author} {\bibfnamefont {R.~E.}\ \bibnamefont {Waltz}},\ }\href
  {https://doi.org/10.1063/1.872666} {\bibfield  {journal} {\bibinfo  {journal}
  {Phys. Plasmas}\ }\textbf {\bibinfo {volume} {5}},\ \bibinfo {pages} {973}
  (\bibinfo {year} {1998})}\BibitemShut {NoStop}%
\bibitem [{\citenamefont {Dimits}\ \emph {et~al.}(2000)\citenamefont {Dimits}
  \emph {et~al.}}]{Dimits2000}%
  \BibitemOpen
  \bibfield  {author} {\bibinfo {author} {\bibfnamefont {A.~M.}\ \bibnamefont
  {Dimits}} \emph {et~al.},\ }\href {https://doi.org/10.1063/1.873896}
  {\bibfield  {journal} {\bibinfo  {journal} {Phys. Plasmas}\ }\textbf
  {\bibinfo {volume} {7}},\ \bibinfo {pages} {969} (\bibinfo {year}
  {2000})}\BibitemShut {NoStop}%
\bibitem [{\citenamefont {Ball}\ \emph {et~al.}(2019)\citenamefont {Ball},
  \citenamefont {Brunner},\ and\ \citenamefont {McMillan}}]{Ball2019}%
  \BibitemOpen
  \bibfield  {author} {\bibinfo {author} {\bibfnamefont {J.}~\bibnamefont
  {Ball}}, \bibinfo {author} {\bibfnamefont {S.}~\bibnamefont {Brunner}},\ and\
  \bibinfo {author} {\bibfnamefont {B.~F.}\ \bibnamefont {McMillan}},\
  }\href@noop {} {\bibfield  {journal} {\bibinfo  {journal} {Plasma Phys.
  Control. Fusion}\ }\textbf {\bibinfo {volume} {61}},\ \bibinfo {pages}
  {064004} (\bibinfo {year} {2019})}\BibitemShut {NoStop}%
\bibitem [{\citenamefont {Vol\v{c}okas}\ \emph {et~al.}(prep)\citenamefont
  {Vol\v{c}okas}, \citenamefont {Ball},\ and\ \citenamefont
  {Brunner}}]{Volcokas2022}%
  \BibitemOpen
  \bibfield  {author} {\bibinfo {author} {\bibfnamefont {A.}~\bibnamefont
  {Vol\v{c}okas}}, \bibinfo {author} {\bibfnamefont {J.}~\bibnamefont {Ball}},\
  and\ \bibinfo {author} {\bibfnamefont {S.}~\bibnamefont {Brunner}},\
  }\href@noop {} {\bibfield  {journal} {\bibinfo  {journal} {Plasma Phys.
  Control. Fusion}\ } (\bibinfo {year} {in prep.})}\BibitemShut {NoStop}%
\bibitem [{\citenamefont {Romanelli}\ and\ \citenamefont
  {Zonca}(1993)}]{Romanelli1993}%
  \BibitemOpen
  \bibfield  {author} {\bibinfo {author} {\bibfnamefont {F.}~\bibnamefont
  {Romanelli}}\ and\ \bibinfo {author} {\bibfnamefont {F.}~\bibnamefont
  {Zonca}},\ }\href {https://doi.org/10.1063/1.860576} {\bibfield  {journal}
  {\bibinfo  {journal} {Phys. Fluids B, Plasma Phys}\ }\textbf {\bibinfo
  {volume} {5}},\ \bibinfo {pages} {4081} (\bibinfo {year} {1993})}\BibitemShut
  {NoStop}%
\bibitem [{\citenamefont {Connor}\ and\ \citenamefont
  {Hastie}(2004)}]{Connor2004}%
  \BibitemOpen
  \bibfield  {author} {\bibinfo {author} {\bibfnamefont {J.~W.}\ \bibnamefont
  {Connor}}\ and\ \bibinfo {author} {\bibfnamefont {R.~J.}\ \bibnamefont
  {Hastie}},\ }\href {https://doi.org/10.1088/0741-3335/46/10/001} {\bibfield
  {journal} {\bibinfo  {journal} {Plasma Phys. Control. Fusion}\ }\textbf
  {\bibinfo {volume} {46}},\ \bibinfo {pages} {1501} (\bibinfo {year}
  {2004})}\BibitemShut {NoStop}%
\bibitem [{\citenamefont {Floquet}(1883)}]{Floquet1883}%
  \BibitemOpen
  \bibfield  {author} {\bibinfo {author} {\bibfnamefont {G.}~\bibnamefont
  {Floquet}},\ }\href@noop {} {\bibfield  {journal} {\bibinfo  {journal} {Ann.
  Sci. de l'Ecole Norm. Superieure}\ }\textbf {\bibinfo {volume} {12}},\
  \bibinfo {pages} {47} (\bibinfo {year} {1883})}\BibitemShut {NoStop}%
\bibitem [{\citenamefont {Bloch}(1929)}]{Bloch1929}%
  \BibitemOpen
  \bibfield  {author} {\bibinfo {author} {\bibfnamefont {F.}~\bibnamefont
  {Bloch}},\ }\href@noop {} {\bibfield  {journal} {\bibinfo  {journal} {Z.
  Physik}\ }\textbf {\bibinfo {volume} {52}},\ \bibinfo {pages} {555–600}
  (\bibinfo {year} {1929})}\BibitemShut {NoStop}%
\bibitem [{\citenamefont {St-Onge}\ \emph {et~al.}(2022)\citenamefont
  {St-Onge}, \citenamefont {Barnes},\ and\ \citenamefont {Parra}}]{StOnge2022}%
  \BibitemOpen
  \bibfield  {author} {\bibinfo {author} {\bibfnamefont {D.~A.}\ \bibnamefont
  {St-Onge}}, \bibinfo {author} {\bibfnamefont {M.}~\bibnamefont {Barnes}},\
  and\ \bibinfo {author} {\bibfnamefont {F.~I.}\ \bibnamefont {Parra}},\
  }\href@noop {} {\bibfield  {journal} {\bibinfo  {journal} {arXiv 2208.02202}\
  } (\bibinfo {year} {2022})}\BibitemShut {NoStop}%
\bibitem [{\citenamefont {Wesson}(2004)}]{Wesson2004}%
  \BibitemOpen
  \bibfield  {author} {\bibinfo {author} {\bibfnamefont {J.}~\bibnamefont
  {Wesson}},\ }\href@noop {} {}\ (\bibinfo  {publisher} {Oxford University
  Press},\ \bibinfo {year} {2004})\BibitemShut {NoStop}%
\bibitem [{\citenamefont {Brakel}\ \emph {et~al.}(2002)\citenamefont {Brakel}
  \emph {et~al.}}]{Brakel2002}%
  \BibitemOpen
  \bibfield  {author} {\bibinfo {author} {\bibfnamefont {F.}~\bibnamefont
  {Brakel}} \emph {et~al.},\ }\href@noop {} {\bibfield  {journal} {\bibinfo
  {journal} {Nucl. Fusion}\ }\textbf {\bibinfo {volume} {42}},\ \bibinfo
  {pages} {903} (\bibinfo {year} {2002})}\BibitemShut {NoStop}%
\end{thebibliography}%

\end{document}